\documentclass[fleqn,usenatbib]{mnras}

\usepackage[T1]{fontenc}
\usepackage{color}
\usepackage{array}

\DeclareRobustCommand{\VAN}[3]{#2}
\let\VANthebibliography\thebibliography
\def\thebibliography{\DeclareRobustCommand{\VAN}[3]{##3}\VANthebibliography}

\usepackage{graphicx}	% Including figure files
\usepackage{amsmath}	% Advanced maths commands
\usepackage{amssymb}	% Extra maths symbols
\usepackage{multirow}

\newcommand{\OIII}{O\,{\scriptsize III}}
\newcommand{\OII}{O\,{\scriptsize II}}
\newcommand{\OI}{O\,{\scriptsize I}}
\newcommand{\CIII}{C\,{\scriptsize III}}
\newcommand{\CII}{C\,{\scriptsize II}}
\newcommand{\CIV}{C\,{\scriptsize IV}}
\usepackage{newtxtext,newtxmath}

\title[Probing Cosmic Dawn with z$\geq$9 galaxies]{Probing Cosmic Dawn : Ages and Star Formation Histories of Candidate $z\geq$9 Galaxies}

\author[N. Laporte et al.]{
N. Laporte,$^{1,2}$\thanks{E-mail: nl408@cam.ac.uk}
R. A. Meyer,$^{3}$
R. S. Ellis, $^{3}$
B. E. Robertson, $^{5}$ 
J. Chisholm, $^{6,7}$
\newauthor
\& 
G. W. Roberts-Borsani $^{4}$
\\
$^{1}$ Kavli Institute for Cosmology, University of Cambridge, Madingley Road, Cambridge CB3 0HA, UK\\
$^{2}$ Cavendish Laboratory, University of Cambridge, 19 JJ Thomson Avenue, Cambridge CB3 0HE, UK\\
$^{3} $Department of Physics and Astronomy, University College London, Gower Street, London WC1E 6BT, UK\\
$^{4}$ Department of Physics and Astronomy, University of California, Los Angeles, 430 Portola Plaza, Los Angeles, CA 90095, USA \\
$^{5}$ University of California–Santa Cruz, 1156 High Street, Santa Cruz, CA 95064, USA \\
$^{6}$  Department of Astronomy, University of Texas, 2515 Speedway, Austin, TX 78712, USA \\
$^{7}$ Hubble Fellow
}

\date{Accepted 2021 April 16. Received 2021 March 5; in original form 2020 November 12}

\pubyear{2020}

\begin{document}
\label{firstpage}
\pagerange{\pageref{firstpage}--\pageref{lastpage}}
\maketitle

\begin{abstract}
We discuss the spectral energy distributions and physical properties of six galaxies whose photometric redshifts suggest they lie beyond a redshift $z\simeq$9. Each was selected on account of a prominent excess seen in the \textit{Spitzer}/IRAC 4.5$\mu$m band which, for a redshift above $z=9.0$,  likely indicates the presence of a rest-frame Balmer break and a stellar component that formed earlier  than a redshift $z\simeq10$. In addition to constraining the earlier star formation activity on the basis of fits using stellar population models with BAGPIPES, we have undertaken the necessary, but challenging, follow-up spectroscopy for each candidate using various combinations of Keck/MOSFIRE, VLT/X-shooter, Gemini/FLAMINGOS2 and ALMA. Based on either Lyman-$\alpha$ or [\OIII] 88 $\mu$m emission, we determine a convincing redshift of $z$=8.78 for GN-z-10-3 and a likely redshift of $z$=9.28 for the lensed galaxy MACS0416-JD. For GN-z9-1, we conclude the case remains promising for a source beyond $z\simeq$9. Together with earlier spectroscopic data for MACS1149-JD1, our analysis of this enlarged sample provides further support for a cosmic star formation history extending beyond redshifts $z\simeq$10. We use our best-fit stellar population models to reconstruct the past rest-frame UV luminosities of our sources and discuss the implications for tracing earlier progenitors of such systems with the \textit{James Webb} Space Telescope.
\end{abstract}


%%%%%%%%%%%%%%%%%%%%%%%%%%%%%%%%%%%%%%%%%%%%%%%%%%

%%%%%%%%%%%%%%%%% BODY OF PAPER %%%%%%%%%%%%%%%%%%

\section{Introduction}
\label{sec.intro}

An important goal of contemporary cosmology is to locate the epoch when the universe was first bathed in starlight, popularly referred to as `cosmic dawn'. According to numerical simulations, dark matter
halos sufficiently massive to induce star formation could develop as early as 150-250 Myr after the Big Bang, corresponding to the redshift range $z\simeq$15-20 \citep{Wise2014,Villanueva2018}.  With some assumptions about their ionising capabilities, studies based on the demographics of early galaxies \citep{Robertson2015} have claimed to be compatible with independent constraints on early reionisation based on analyses of the optical depth of electron scattering to the cosmic microwave background (CMB; \citealt{Planck2018}), as well as tentative measures of 21cm absorption at a redshift $z\simeq$16-19 in the CMB possibly associated with UV photons produced by early star formation \citep{Bowman2018}.

Recent deep imaging with the \textit{Hubble} and \textit{Spitzer} Space Telescopes has extended our redshift horizon to $z\simeq$9-11 \citep{Ellis13,Oesch2018}. At this redshift, the universe is less than 600 Myr old and the presence of main sequence stars older than 250 Myr would imply galaxy formation originating before a redshift $z\simeq$14. Such a case of `age-dating' a high redshift galaxy was presented by \citet{Zheng2012} for the lensed galaxy MACS1149-JD1 based upon a prominent photometric excess seen in the \textit{Spitzer}/IRAC 4.5$\mu$m band. For the assumed photometric redshift of $z$=9.6 $\pm$ 0.2, the IRAC excess would imply the presence of a Balmer break indicative of a burst of star formation some 200-300 Myr earlier. However, central to this conclusion is excluding the possibility of a redshift below $z\simeq$9 where such an IRAC excess could arise from contamination of the 4.5$\mu$m band by intense rest-frame optical [\OIII] 5007 \AA\ emission consistent with a much younger stellar population \citep{Labbe13,Roberts-Borsani2016}. 
% An early attempt to secure a spectroscopic redshift of MACS1149-JD1 using a grism onboard the \textit{Hubble Space Telescope} (HST) \citep{Hoag2018} 
An early attempt to secure a spectroscopic redshift of MACS1149-JD1 using grism capabilities onboard the \textit{Hubble} Space Telescope (HST) \citep{Hoag2018} provided a tantalising hint of a Lyman-break at $z>9$, however the lack of an emission line prevented a secure redshift constraint. A convincing redshift of $z=9.11$ was eventually determined through Lyman-$\alpha$ and [\OIII] 88 micron emission by \citet{Hashimoto2018}, using ALMA and VLT/X-Shooter. The subsequent re-analysis of the object using the secure spectroscopic redshift and improved IRAC photometry resulted in an estimated age of 290 $\pm$ 150 Myr for the dominant stellar component, consistent with a formation redshift of 15.4 $\pm$2.3.

Various studies have discussed whether MACS1149-JD1 might be representative of the galaxy population at early times and hence provide a realistic first estimate of when cosmic dawn occurred. However, accurate interpretation of the associated photometry is not without its challenges. For instance, although \citet{Hashimoto2018} derived a stellar mass of $\simeq 10^9 M_{\odot}$ for MACS1149-JD1, this relied on an assumed and uncertain lensing magnification, thus introducing important uncertainties when comparing its properties to theoretical predictions. Furthermore, a fundamental difficulty has been that a prominent Balmer break requires a {\em declining} star formation rate with cosmic time for at least some portion of the early history of a galaxy, something that is difficult to reconcile with contemporary numerical simulations \citep{Binggeli2019}. 

Observationally, several $7.5<z<9$ objects (both lensed and unlensed) with secure spectroscopic redshifts display a significant IRAC excess which could be indicative of the presence of a Balmer break and early episodes of star formation (e.g., \citealt{Tamura2019},\citealt{Bakx2020}, \citealt{Strait2020}). However, here the interpretation is confused by the likely prominence of [\OIII] 5007 \AA\ emission. To break the degeneracy of line emission and starlight as contributors to the IRAC excess, \citet{Roberts-Borsani2020} combined HST and \textit{Spitzer}/IRAC photometry with ALMA far-infrared [\OIII] 88 $\mu$m emission and dust continuum constraints for four $z>7$ sources (including MACS1149-JD1). Although their analysis gave a lower age for MACS1149-JD1 than \citet{Hashimoto2018}, they nonetheless concluded that, collectively for these four sources, nearly half the stellar mass was produced prior to $z\simeq$10. 

 Even for sources believed to be beyond $z\simeq9$, it is possible that an IRAC excess may not entirely arise via a Balmer break. For MACS1149-JD1 \citet{Hashimoto2018} ruled out strong contamination of the 4.5$\mu$m band from intense [O III] 4959 \AA\ and H$\beta$, which would have indicated a much younger system. Likewise, very strong [O II] 3727 \AA\ emission might contaminate the band, but only if the true redshift approaches $z\simeq$10. Using hydrodynamical simulations and a comoving volume of 70 Mpc$^3$, \citet{Katz2019} predicted the presence of three massive $z>9$ galaxies with similarly strong IRAC colours and matched photometry to MACS1149-JD1. While their results also point to likely episodes of extended star formation, they conclude that dust may contribute to the IRAC excess, thereby reducing the strength of the Balmer break.

Given the implications for the birth of galaxies, determining the amount of star formation prior to $z\simeq$9 is an important, but challenging endeavour. While ambiguities of interpretation may remain given the signal/noise of the available photometric data, our motivation in this paper is to further explore the Balmer break interpretation for six promising $z\geq$9 candidates (including MACS1149-JD1) each with a 4.5$\mu$m IRAC excess. A major advance is a concerted attempt to secure their spectroscopic redshifts. Our goal is to explore the possible extent of star formation activity prior to a redshift $z\simeq$10. We consider this timely given it may have important consequences for exploring this early period further with the {\it James Webb} Space Telescope.

A plan of the paper follows. In Section \ref{sec.selection}, we describe the selection criteria we applied to construct our sample of $z\geq$9 objects. In Section \ref{sec.photometry} we present the relevant spectral energy distributions (SEDs) and use these to infer the likely photometric redshifts and their star formation histories. This analysis provides a first assessment of the early star formation history for an enlarged population of $z\simeq$9 galaxies. The results of our spectroscopic follow-up campaign are described in Section \ref{sec.spectro}. Given the challenges of working at these limits, we are not able to convincingly secure a redshift for all of our candidates.  Section \ref{sec.prop} returns to a discussion of the physical properties of the sample in the light of the limited spectroscopic data. We also discuss the extent to which our IRAC excess sample may be representative of the broader population of $z\simeq$9 galaxies and the frequency with which Lyman-$\alpha$ emission is being located deep in the reionisation era. Finally we discuss the implication of our findings in the context of  direct searches for earlier activity with the \textit{James Webb} Space Telescope in Section \ref{sec.discussion}. 

Throughout this paper we use a concordance cosmology with $H_0 = 70, \Omega_M = 0.3, \Omega_L = 0.7$. Magnitudes are in the AB system \citep{AB}. 

\section{Selection of z>9 sources}
\label{sec.selection}

To make further progress in charting the cosmic star formation history beyond the limits directly accessible to HST,  we have selected further candidates for detailed study from three near-infrared surveys: CANDELS \citep{CANDELS} and the Frontier Fields \citep{Lotz2017} from \textit{Hubble} Space Telescope imaging, and the UltraVISTA ground-based campaign \citep{UltraVISTA}. 

We applied the following selection criteria to likely $z\simeq$9 candidates with $m_{\text{H-band}}<26.5$ mag (see Section \ref{sec.photometry}): (i) a 2$\sigma$ non-detection in all bands bluewards of the Lyman-break at the putative redshift; (ii) a 5$\sigma$ detection in two consecutive bands redwards of the Lyman-break; (iii) red \textit{Spitzer}/IRAC colours (3.6$\mu$m - 4.5$\mu$m > 0.5); (iv) a photometric redshift probability distribution permitting $z\geq$9 at 1$\sigma$.  We used publicly available catalogues for the CANDELS and UltraVISTA fields. AstroDeep catalogues were used for four Frontier Field clusters (A2744, MACS0416, MACS0717 and MACS1149) and we constructed our own catalogues for the clusters A370 and AS1063 following the method described in \citet{Laporte2016}. The \textit{Spitzer}/IRAC data used in this paper are drawn from several surveys and therefore not uniform in depth. CANDELS IRAC data have a 5$\sigma$ depth of 25.5 mag; the 5$\sigma$ depth of the UltraVISTA IRAC data is 25.8 mag and the Frontier Fields data reach a 5$\sigma$ depth of 26.5 mag. We extracted the Spitzer photometry using a 1.2 arcsec radius aperture centred at the position of the source on the F160W image, and applied the usual aperture correction. However noting that, for blended sources, this may overestimate the IRAC flux, in those cases we extracted the photometry more carefully, taking into account the shape of the galaxies with GALFIT \citep{Peng2010} assuming a Sersic profile.

Including MACS1149-JD1, these criteria led to six suitable sources distributed as follows : two (presumably lensed) sources viewed behind a Frontier Fields cluster, three in the CANDELS fields GOODS-N and GOODS-S, and one in UltraVISTA survey (see Table \ref{list}). The most restrictive criterion in their selection is that relating to the red IRAC colour as this can only be applied to relatively bright targets given the depth difference between $\sim$1$\mu$m data and the IRAC images. 
All these sources have already been identified as promising high-redshift candidates by previous studies \citep{Oesch2018,Ishigaki2018,Bowler2020}.

%%%%%%%%%%%%%%%%%%%%%%%%%%%%%%%%%%%%%%%%%%%%%%
\begin{table*}
\hspace{-1.4cm} 
    \centering    

   \hspace{-1.4cm}  \begin{tabular}{l|ccccccc} \hline
Target & RA & DEC & $z_{\text{phot}}$ & $\rm{M_{uv}}$ & $\rm{M_{\star}}$[$\times$10$^9$M$_{\odot}$] &  Age (Myr) & $f_M$(z$>$10) [\%]\\ \hline
MACS0416-JD$^a$ & 04:16:11.52 & -24:04:54.0 & 9.25$^{+0.08}_{-0.09}$ & -20.83$\pm$0.22 & 1.50$^{+0.84}_{-0.61}$ &  360$^{+108}_{-157}$  & 74.1$^{+8.1}_{-25.6}$\\
\multirow{2}{*}{MACS1149-JD1$^b$}& \multirow{2}{*}{11:49:33.59} & \multirow{2}{*}{22:24:45.76} & 9.44$^{+0.02}_{-0.03}$ & -19.17$\pm$0.04 & 0.44$^{+0.05}_{-0.04}$  & 484$^{+17}_{-36}$  & 86.5$^{+2.0}_{-4.3}$ \\
  &   &   & \textit{9.11} & -19.12$\pm$0.04 & 0.66$^{+0.09}_{-0.04}$ &   192$^{+159}_{-87}$  & 47.5$^{+20.0}_{-14.2}$  \\
GN-z10-3$^c$ & 12:36:04.09 & +62:14:29.6 & 9.57$^{+0.23}_{-0.27}$ & -20.72$\pm$ 0.12 & 1.57$^{+0.81}_{-0.63}$ &  265$^{+153}_{-145}$  & 38.0$^{+23.1}_{-38.0}$\\
GN-z9-1$^d$ & 12:36:52.25 & +62:18:42.4 & 9.22$^{+0.32}_{-0.33}$ & -20.81$\pm$ 0.18 & 2.18$^{+1.24}_{-0.85}$ &   323$^{+137}_{-168}$ & 65.1$^{+18.7}_{-32.2}$ \\
GS-z9-1$^d$ & 03:32:32.05 & -27:50:41.7 & 9.26$^{+0.41}_{-0.42}$ & -20.38$\pm$0.20 & 2.47$^{+1.62}_{-1.03}$ &   326$^{+128}_{-176}$  & 70.6$^{+12.5}_{-25.4}$ \\ 
UVISTA-1212$^e$ & 10:02:31.81 & 02:31:17.10 & 8.88$^{+0.27}_{-0.46}$ & -22.93$\pm$0.20 & 9.7$^{+5.10}_{-4.72}$ &  280$^{+172}_{-174}$ & 32.4$^{+47.2}_{-32.4}$ \\ \hline
\end{tabular}
\caption{\label{list} Bright $z_{\text{phot}}\geq$9 galaxies with red IRAC colours (3.6$\mu$m - 4.5$\mu$m > 0.5) with physical properties determined using BAGPIPES \citep{BAGPIPES}. Photometric redshift uncertainties represent 1$\sigma$ values and absolute magnitudes and stellar masses are corrected for lensing magnification where appropriate (see text). The final column displays the inferred percentage of the presently-observed stellar mass formed before $z=$10. \\
$^a$ \citet{Laporte2016}, $^b$ \citet{Zheng2012}, $^c$\citet{Oesch2014}, $^d$\citet{Oesch2018}, $^e$\citet{Bowler2020}
} 
\end{table*}

\section{Photometric Analysis}
\label{sec.photometry}

\begin{figure*}
    \centering
   \includegraphics[width=15cm]{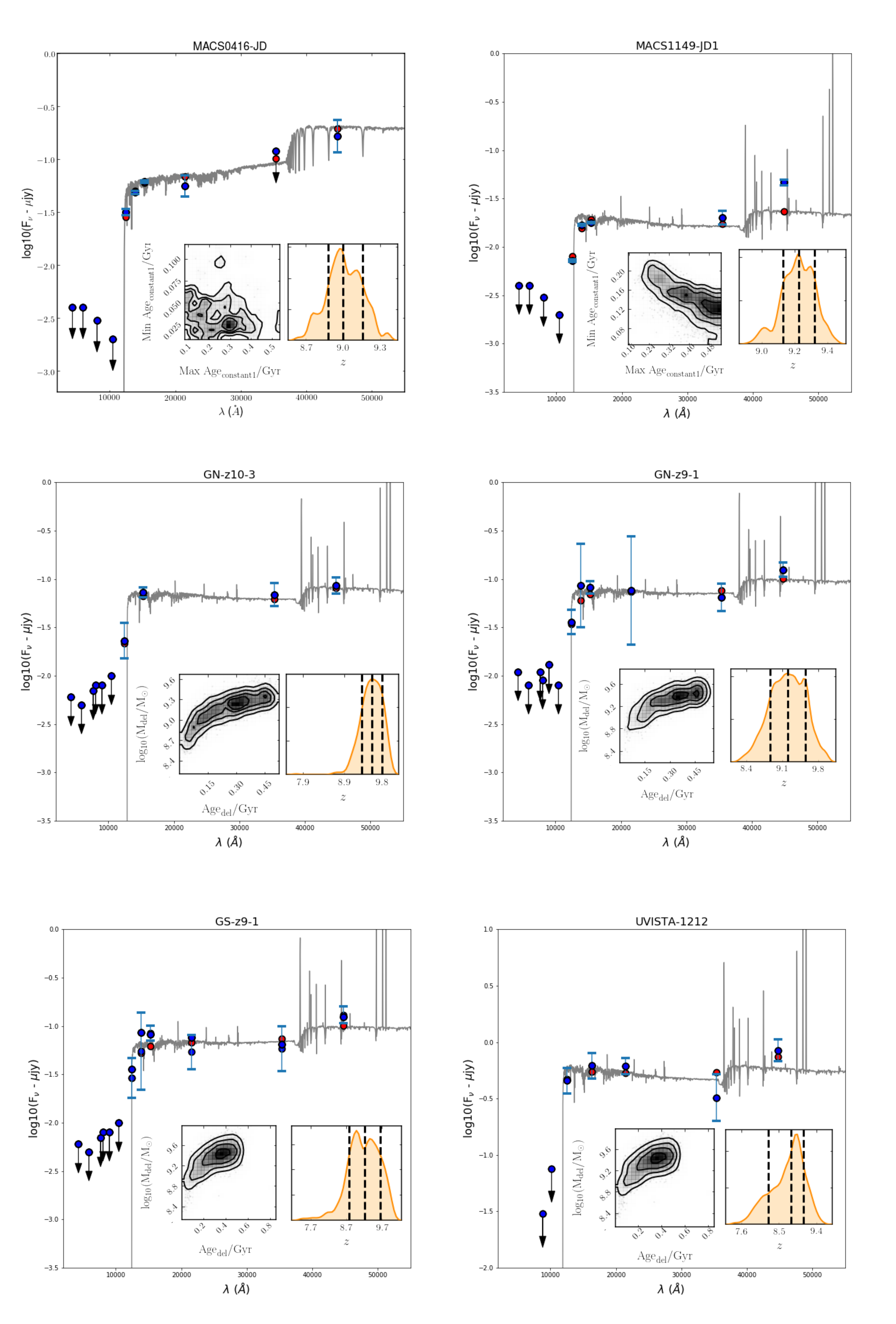} \caption{\label{fig.sed} Spectral energy distributions and best-fit BAGPIPES \citep{BAGPIPES} models for the 6 galaxies in Table \ref{list}. Blue points represent photometric data or limits, and red points represent the predicted fluxes. All models incorporate contributions from nebular emission (see text). The right inset panel shows the redshift probability distribution, while the left inset panel shows the posterior distribution of the stellar mass as a function of the age of the stellar population. HST upper limits are showed at 1$\sigma$, IRAC limits are at 2$\sigma$}.
    \label{fig:M0416-JD_prop}
\end{figure*}

The spectral energy distributions (SEDs) and photometric redshift likelihood distributions for our selected six $z>9$ candidates are presented in Figure \ref{fig.sed}. The SEDs utilise photometry drawn from publicly-available imaging data taken with \textit{Hubble} and \textit{Spitzer} Space Telescopes and, where available, the VISTA deep surveys. We matched the Point-Spread Function (PSF) of HST images using \textit{TinyTim} models \citep{TinyTim}. We extracted the total SED following the method described in \citet{Finkelstein2013} with an aperture correction computed from the \textit{Sextractor} MAG\_AUTO measured on the F160W image from WFC3/IR or the H-band from VISTA. The IRAC/\textit{Spitzer} photometry following method described in the previous section . Two sources in our sample, MACS0416-JD and MACS1149-JD1, are located behind Frontier Fields clusters and thus are  gravitationally-lensed. We estimated the magnification for these using the online magnification calculator for the Frontier Field
survey\footnote{https://archive.stsci.edu/prepds/frontier/lensmodels/\#magcalc} which is based on lensing models developed by \citet{Hoag2016}, \citet{Caminha2017}, \citet{Jauzac2014}, \citet{Richard2014}, \citet{Kawamata2016}, \citet{Kawamata2018}, \citet{Ishigaki2015}, \citet{Johnson2014} and \citet{Grillo2015}. This yields a magnification of 1.90$^{+0.03}_{-0.02}$ and 28$^{+44}_{-11}$ respectively for MACS0416-JD and MACS1149-JD1. The latter estimate represents a revision of the magnification adopted by \citet{Hashimoto2018}.

We simultaneously determine the optimal photometric redshift and a range of physical properties for the six selected galaxies using BAGPIPES \citep{BAGPIPES}. We allow the photometric redshift to lie in the range $0.0<z_{\text{phot}}<10.0$ which, for $z\leq9.0$, permits the IRAC excess to be formed, in part, from [\OIII] line emission \citep{Roberts-Borsani2020}. BAGPIPES generates HII regions from CLOUDY photoionisation code \citep{CLOUDY} and assumes that the nebular emission is the sum of emission from HII regions of different ages following \citet{Byler2017}. In particular, it is capable of reproducing intense [\OIII] emission with equivalent widths EW[\OIII+H$\beta$]$>$1000\AA\ , sufficient for consideration of dominant contributions to an IRAC excess at high redshift. BAGPIPES assumes that the gas-phase metallicity is similar to that of the stars producing the ionising radiation (with a range of $Z\in$[0.0:2.5] $Z_{\odot}$). Finally, BAGPIPES accounts for dust emission via the \citet{Draine2007} model, with a reddening law following the \citet{Calzetti2001} relation. We fit the SED of our candidates assuming a single-component model with four possible star-formation histories (SFH) : delayed, exponential, constant or burst-like. The best SED-fit is selected to be that with the largest likelihood and is presented as the grey curve for each candidate in Figure \ref{fig.sed}, with the associated physical parameters listed in Table \ref{list}. The best fit is always obtained for the delayed or constant SFH. Although we tabulate the age of the best-fit stellar population, a more meaningful measure is the fraction (expressed as a percentage) of the observed stellar mass that was assembled prior to a redshift $z=10$. For the two gravitationally-lensed galaxies, the luminosities and stellar masses have been corrected for the magnifications assumed above. Table~\ref{tab.fit} shows the range of parameters adopted in BAGPIPES for each particular SFH.  

\begin{table}
\centering    
   \begin{tabular}{|c|c|c|c|c|} 
   \hline
   Parameters & Burst & Exponential & Constant & Delayed \\ \hline
   t$_{\rm SF\ begin}$ [Gyr]& 0.0-0.1 & \multicolumn{3}{c|}{0.0 ; 1.0} \\ \hline
   t$_{\rm SF\ end}$ [Gyr] & - & - & 0.0 ; 1.0 & - \\ \hline
   $\tau$ [Gyr] & - & 0.1 ; 10.0 & - & 0.1 ; 10.0 \\ \hline
   $\log$M$_{\star}$ [M$_{\odot}$] & \multicolumn{4}{c|}{0.0 ; 12.0} \\ \hline
    Z [$Z_{\odot}$] & \multicolumn{4}{c|}{0.0 ; 2.5} \\ \hline
    Av [mag] & \multicolumn{4}{c|}{0.0 ; 3.0} \\ \hline
    log U & \multicolumn{4}{c|}{-4.0 ; 0.0} \\ \hline
    \end{tabular}
    \caption{\label{tab.fit} The BAGPIPES parameter ranges used to fit the SEDs of the 6 targets studied in this paper. The age (or maximum age) is chosen to be less than 1 Gyr given the selection criteria applied to build our sample. The ionisation parameter range we used is large to account for extreme values ($\log$ U$>$-2) as observed in previous high-redshift galaxies \citep{Stark2017}.} 
\end{table}

Despite the different HST and ground-based datasets from which our six IRAC excess sources were selected, they span a limited range in luminosity and stellar mass. As a consequence of its detection to the shallower magnitude limit with a ground-based telescope, UVISTA-1212 is the most luminous and massive galaxy in the sample. Later in Section \ref{sec.prop} we discuss how representative is our sample with respect to the larger population of sources thought to be at $z\simeq$9. For all of the sources except UVISTA-1212, the most probable photometric redshift derived is above $z\simeq$9 and thus a Balmer break may be the most likely explanation for the IRAC excess. Nonetheless, as discussed by \citet{Hashimoto2018} and \citet{Roberts-Borsani2020}, alternative explanations based on contributions from strong nebular emission lines or dust reddening can be considered.

 Although [\OIII] 5007 \AA\ is redshifted out of the IRAC 4.5$\mu$m band for $z>9.1$, \citet{Hashimoto2018} considered the contribution of H$\beta$ and [\OIII] 4959\AA . Even for a very young starburst with a hard radiation field (log U = -2.0) and low metallicity (20\% solar), it is not possible to reproduce a significant IRAC excess. Furthermore, for the youngest ages, the IRAC 3.6$\mu$m flux is raised by the presence of a significant nebular continuum. Likewise, [\OII] 3727 \AA\ will enter the 4.5$\mu$m band for redshifts above $z\simeq$9.7. The photometric likelihood distribution does extend to such a high redshift for some of our sources (Figure \ref{fig:M0416-JD_prop}). However, the IRAC excess would only be matched if [\OII] had a rest-frame equivalent width of over 1000 \AA\ which seems improbable. Finally, it is possible to match the SEDs by combining a delayed SFH with dust extinction. In this case, to match the IRAC excess, the extinction would typically need to exceed $A_V>$1.7 mag, which is larger than that inferred in previous measures of reionisation era galaxies (e.g. \citealt{Schaerer2010}, \citealt{Strait2020}).

In Figure \ref{fig:M0416-JD_prop} we show the posterior distribution of the age and stellar mass to indicate the reliability of concluding there was star formation prior to $z=10$ as listed in Table \ref{list}. Despite uncertainties in the available photometry, the shape of the posterior is reasonably well-defined for ages greater than 200 Myr, indicating the likelihood of significant earlier star formation. The BAGPIPES solution for MACS1149-JD1 can be compared with the solutions derived by \cite{Hashimoto2018} and \citet{Roberts-Borsani2020}, both of which included ALMA [\OIII] 88$\mu$m fluxes and upper limits on the dust continuum in Band 7 as additional constraints. Although Roberts-Borsani et al advocated the use of ALMA to break the degeneracy between optical [\OIII] emission and a Balmer break as contributors to the IRAC excess, it is not a key issue in this case given the spectroscopic redshift of $z$=9.11 precludes the former case. Although an academic exercise given its redshift is known, both our photometric redshift ($z_{\text{phot}}=9.44^{+0.02}_{-0.03} $) and age ( $484 ^{+17}_{-36}$ Myr) are consistent with the range of early estimates. We include a separate entry in Table \ref{list} for the case where the redshift is fixed at its spectroscopic value of $z$=9.11 \citep{Hashimoto2018}. %As noted by Roberts-Borsani et al, the age is then reduced (here to $484 \pm 40$ Myr) because, at the lower redshift, the Balmer break lies entirely shortward of the IRAC 4.5$\mu$m filter.

We emphasise that the inferred contribution of early star formation is uncertain in many cases, particularly for GN-z10-3 whose IRAC excess is the least prominent. Nonetheless, overall our sample is consistent with having significant star formation prior to $z\simeq$10, as was concluded for MACS1149-JD1.  We will return to a further discussion of this in Section \ref{sec.discussion}.

%\begin{table}
%\hspace{-1.4cm} 
%    \centering    

%   \hspace{-1.4cm}  \begin{tabular}{l|cccc} \hline
%Target & SFH & $z_{\text{phot}}$ & $z_{form}$ & M$_{\star}$ \\
%        &   &       &       &   $\times$10$^9$M$_{\odot}$ \\ \hline
%GN-z10-3 & Burst & 9.59$\pm$0.25 & 10.1$\pm$0.1 & 1.68$\pm$0.73 \\
%        & \textit{Combination} & 9.59$\pm$0.27 & 10.0$^{+10.5}_{-4.0}$ & 1.29$\pm$1.28 \\

%GN-z9-1 & Combination & 9.23$\pm$0.3 & 13.6$^{+6.7}_{-3.1}$ & 2.45$\pm$2.02\\
%MACS0416-JD & Combination & 9.04$\pm$0.13 & 13.1$^{+5.1}_{-2.6}$ & 6.3$\pm$2.4 \\
%MACS1149-JD1 & Burst & 9.37$\pm$0.05 & 11.05$^{+0.75}_{-0.67}$ & 1.92$\pm$0.38 \\
%        & \textit{Combination} & 9.35$\pm$0.05 & 16.7$^{+13.3}_{-4.6}$ & 1.88$\pm$0.37 \\
%UVISTA-1212 & Burst & 8.77$\pm$0.41 & 8.9$\pm$0.4 & 6.19$\pm$2.72 \\
%        & \textit{Combination} & 8.91$\pm$0.30 & 10.8$\pm$0.9 & 16.6$\pm$7.4 \\
        
%GS-z9-1 & Combination & 9.29$\pm$0.39 & 14.3$^{+6.5}_{-3.3}$ & 2.9$\pm$2.6 \\ \hline

%    \end{tabular}
%    \caption{\label{sed.prop} Physical properties deduced from SED-fitting using BAGPIPES \citep{BAGPIPES}. The second column shows the Star Formation History (SFH) for which the best fit is obtained : a single Burst, a Constant Star Formation episode, an exponentially declining star formation and a combination of a single burst with an episode of constant star formation }
%\end{table}

\section{Spectroscopic Follow-up}
\label{sec.spectro}
Although Lyman-$\alpha$ is the most prominent rest-frame UV emission line in star-forming galaxies at intermediate redshift \citep{Stark2010,Stark2011}, its visibility deep in the reionisation era at $z\simeq$9 is expected to be significantly diminished via resonant scattering by neutral hydrogen. For this reason, various efforts have been made to target metallic UV lines such as [\CIII]1909 \AA\ and \CIV 1548 \AA\ with limited success \citep{Laporte2017,Mainali2020}. Surprisingly, for 11 galaxies with spectroscopic redshifts above $z=7.5$, Lyman-$\alpha$ was detected in 7 with additional UV metallic lines in only 4 cases. In all cases, detecting the intrinsically fainter metallic lines has been more challenging than searching for Lyman-$\alpha$.

Far infrared emission lines, such as [\OIII] 88 $\mu$m arising from ionised gas and [\CII] 158 $\mu$m from the neutral gas, have been pivotal in spectroscopic confirmation at high redshifts with the ALMA interferometer \citep{Inoue2014}. For the same 11 $z>7.5$ galaxies of which 7 are accessible with ALMA, [\OIII] and/or [\CII] emission have been detected in 5 cases. Moreover, in cases where both ALMA and near-infrared spectrographs on ground-based telescopes were both used in similar exposure times (e.g., \citealt{Laporte2017}, \citealt{ Hashimoto2018}), the ALMA detections were more significant. For this reason, we adopted a multi-facility approach in attempting to secure spectroscopic redshifts for the targets introduced in Table \ref{list}.

We discuss the results of our spectroscopic campaigns for each target in turn below. A summary of the results of our spectroscopic campaign is given in Table 2.

\subsection{MACS0416-JD}
\label{sec.macs0416}

\begin{figure}
    \centering
        \includegraphics[width=8cm]{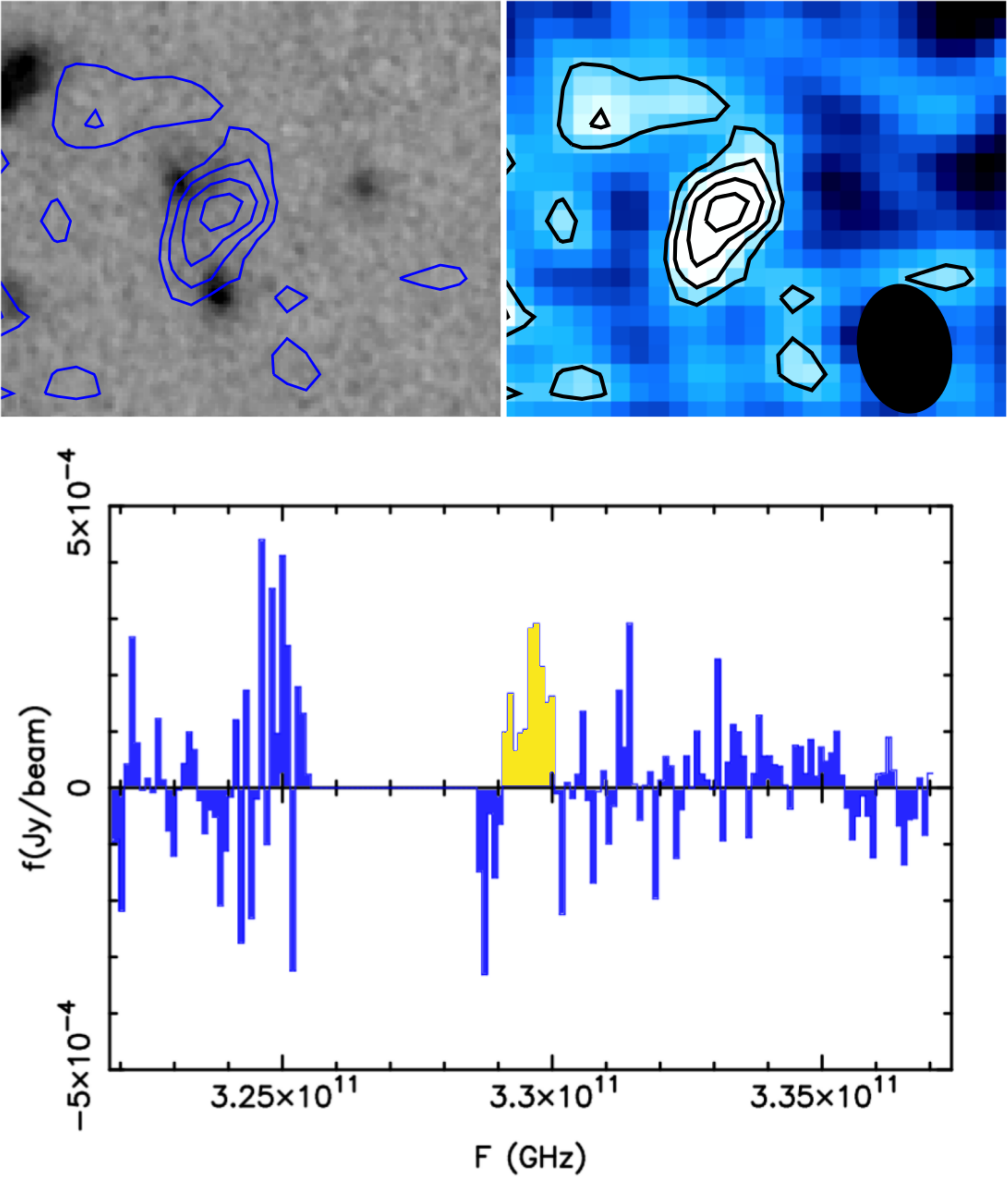}
        \smallskip
        \includegraphics[width=9cm]{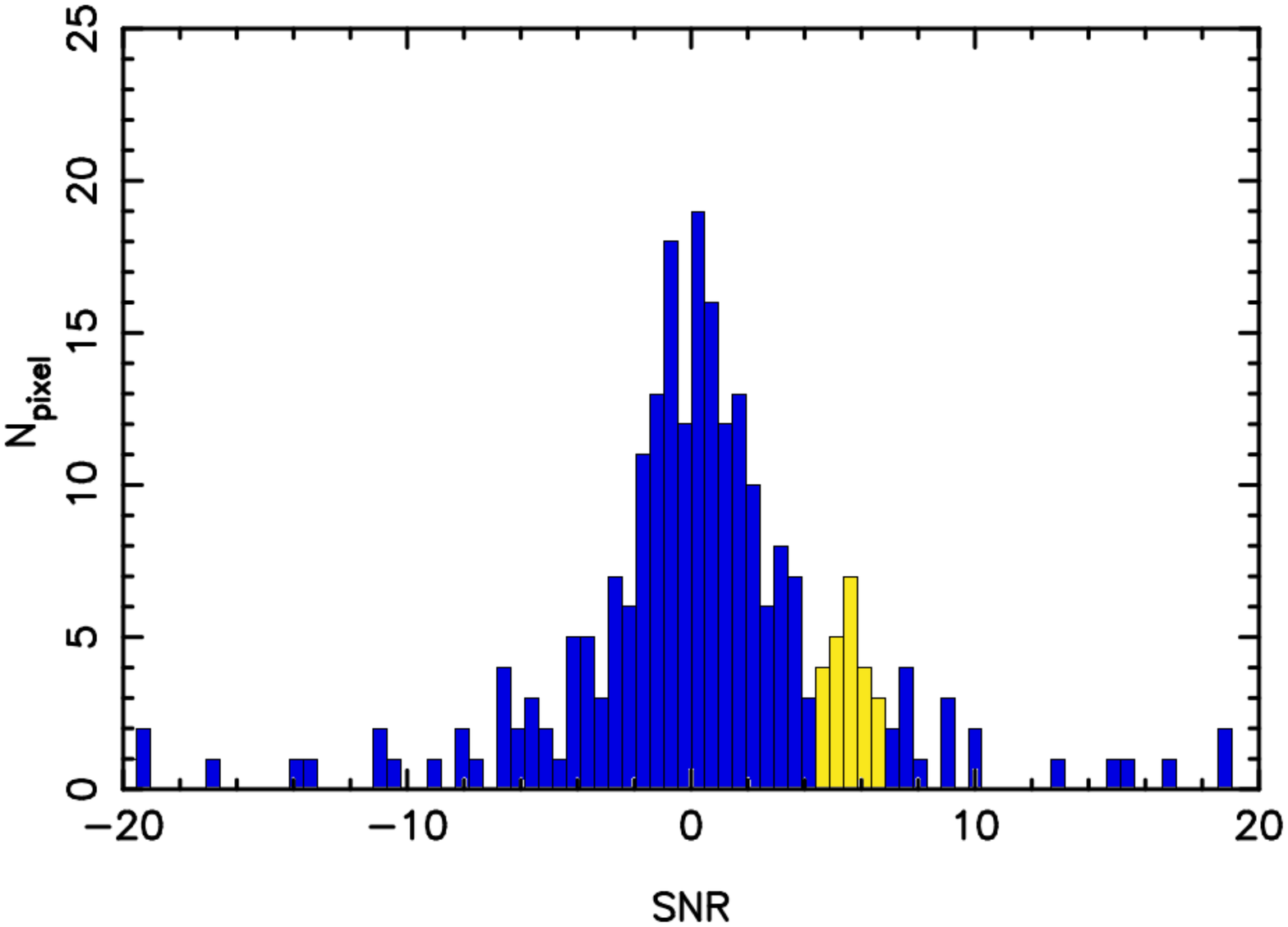} 
    \caption{\label{MACS0416-JD} (\textit{Top}) ALMA contours overplotted on the HST F160W image (left) and ALMA band 7 data (right) from a 2$\sigma$ level upwards for MACS0416-JD. The beam size is indicated at the bottom right of the ALMA panel. (\textit{Centre}) Extracted 1D spectrum over the full frequency range sampled showing a clear [\OIII] 88 $\mu$m emission line at 329.69 GHz corresponding to a redshift of $z$=9.28. (\textit{Bottom})
    Distribution of the pixel signal/noise ratio (SNR) along the line of sight at the position of MACS0416-JD. In the case of non-detection, the histogram shape should be consistent with a gaussian. The yellow region highlights a deviation from a gaussian shape at SNR$\simeq$6.  }
\end{figure}

We obtained 12.6hrs (8hrs on source) ALMA band 7 observation in Cycle 6 (ID: 2019.1.00061.S, PI: R. Ellis) to search for [\OIII] 88 $\mu$m in MACS0416-JD. The spectral windows were setup to cover the 1-$\sigma$ redshift probability distribution estimated from SED-fitting (Figure \ref{fig.sed}, viz. $\nu$ $\in$ [321.8 GHz : 325.7 GHz], $\cup$ [328.6 GHz : 335.6 GHz ], and $\cup$ [340.6 GHz : 347.6 GHz]. The data were obtained between October 2019 and January 2020 and reduced using the ALMA pipeline (v. 5.6.1-8) with a natural weighting scheme, a uv-tapper of 0.35 arcsec with a channel width of 16 MHz and a beam size of 1.00$\times$0.76 arcsec in order to maximise the signal to noise ratio of any emission line (\citealt{Tamura2019}). At the HST position of MACS0416-JD we identified a clear feature at $\nu$=329.69 GHz corresponding to a redshift $z$=9.28 (Figure \ref{MACS0416-JD}). 

The offset between the centroid of the UV continuum and the peak of [OIII]88$\mu$m emission is 0.40 arcsec, corresponding to an observed physical separation of $\simeq$1.8 kpc at $z\sim$9.28. Such a small displacement is consistent with previous [OIII]88$\mu$m observations at high-$z$ (e.g. \citealt{Carniani2017}, \citealt{Harikane2020}).  We measured the integrated flux on a map created using the \textit{immoments} CASA recipe. We determine $f_{int}$=0.298$\pm$0.051 Jy.km.sec$^{-1}$ which is within the range values found previously in $z\geq$8 galaxies (\citealt{Tamura2019}, \citealt{Hashimoto2018}). Assuming a Gaussian profile, we measure a FWHM of 313$\pm$22 km sec$^{-1}$. The significance of the line can be estimated via a histogram of the pixel by pixel signal to noise ratio (SNR) at the spatial position of the target. In the absence of emission, this will approximate a Gaussian distribution. If  emission is present, a second peak will be seen indicating its significance. The lower panel of Figure \ref{MACS0416-JD} reveals such a second peak at  SNR$\simeq$6 consistent with the analysis above. No dust continuum was detected to a 1-$\sigma$ level of 18$\mu$Jy/beam.

\begin{figure}
    \centering
        \includegraphics[width=9cm]{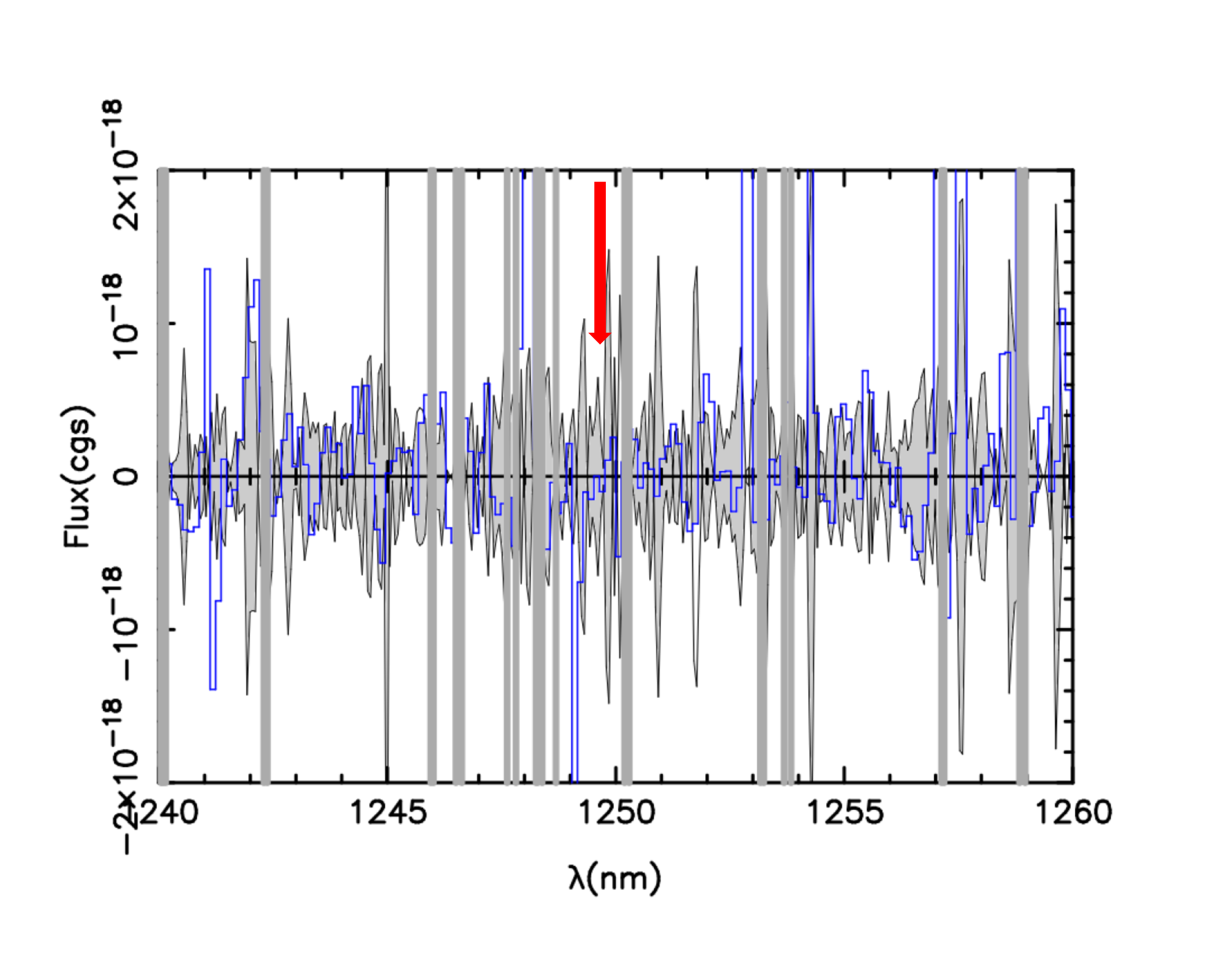}
    \caption{\label{fig.MACS0416-JD_X-Shooter} Portion of the X-Shooter spectra of MACS0416-JD indicating no convincing detection of Lyman-$\alpha$ at the redshift of [\OIII] 88 $\mu$m (red arrow). Grey rectangles show the position of sky lines. The grey line shows the 1$\sigma$ noise spectrum and blue line shows the extracted spectrum at the position of MACS0416-JD. }
\end{figure}

We also observed this target in service mode with X-Shooter/VLT between August and September 2019 (ID: 0104.A-0028(A), P.I. : R. Ellis). The target was centred using a small blind offset ($<$20 arcsec) in the 0.9 arcsec JH slit to reduce the background and a nod length of 4 arcsec was used. The total on source exposure time in the near-infrared arm was 10hrs in excellent seeing condition ($<$0.8 arcsec). The data were reduced using the latest version of the X-shooter pipeline (3.5.0) and inspected visually by two authors (NL and RAM). No emission line was found in any of the three X-Shooter arms. Given the ALMA-based redshift, the absence of Lyman-$\alpha$ implies a 1$\sigma$ upper limit of 1.8$\times$10$^{-18}$cgs assuming a FWHM=200km/s (Figure \ref{fig.MACS0416-JD_X-Shooter}).

\begin{table*}
    \centering 
    \hspace{0.5cm}
   { \small
   \hspace{-1.6cm}  
\begin{tabular}{l|cc|c|c|}
Target & Redshift & Instruments & Feature & Flux / 1$\sigma$  \\ \hline
MACS1149-JD1$^{\star}$ &  9.11 & X-SHOOTER & Ly-$\alpha$ & (4.3$\pm$1.1)$\times$10$^{-18}$cgs \\
MACS0416-JD & 9.28 & X-SHOOTER & Ly-$\alpha$ & $<$1.8$\times$10$^{-18}$cgs \\
GN-z10-3 & 8.78 & MOSFIRE & Ly-$\alpha$ & (1.37$\pm$0.24)$\times$10$^{-18}$cgs \\ 
GN-z9-1 & \textit{9.89?} & MOSFIRE & Ly-$\alpha$ &  (2.2$\pm$0.7)$\times$10$^{-18}$cgs \\ 
U-1212 & - & FLAMINGOS2 & Ly-$\alpha$ & $<$4$\times$10$^{-18}$ cgs\\ 
GS-z9-1 & - & X-SHOOTER & Ly-$\alpha$ & $<$ 7$\times$10$^{-19}$ cgs \\ \hline 
MACS0416-JD & 9.28 & ALMA & [\OIII] 88 $\mu$m & 0.65$\pm$0.13 mJy/beam \\  
MACS1149-JD1  & 9.11 & ALMA & [\OIII] 88 $\mu$m & 0.84$\pm$0.11 mJy/beam \\ \hline

\end{tabular}
}
    \caption{\label{tab.spectro} Spectroscopic observations of $z\geq$9 candidates. For the [\OIII] 88 $\mu$m line, we report the peak flux. \quad $^{\star}$ from \citet{Hashimoto2018}}
\end{table*}

\subsection{GN-z10-3 and GN-z9-1}

\begin{figure}
    \centering
        \includegraphics[width=8cm]{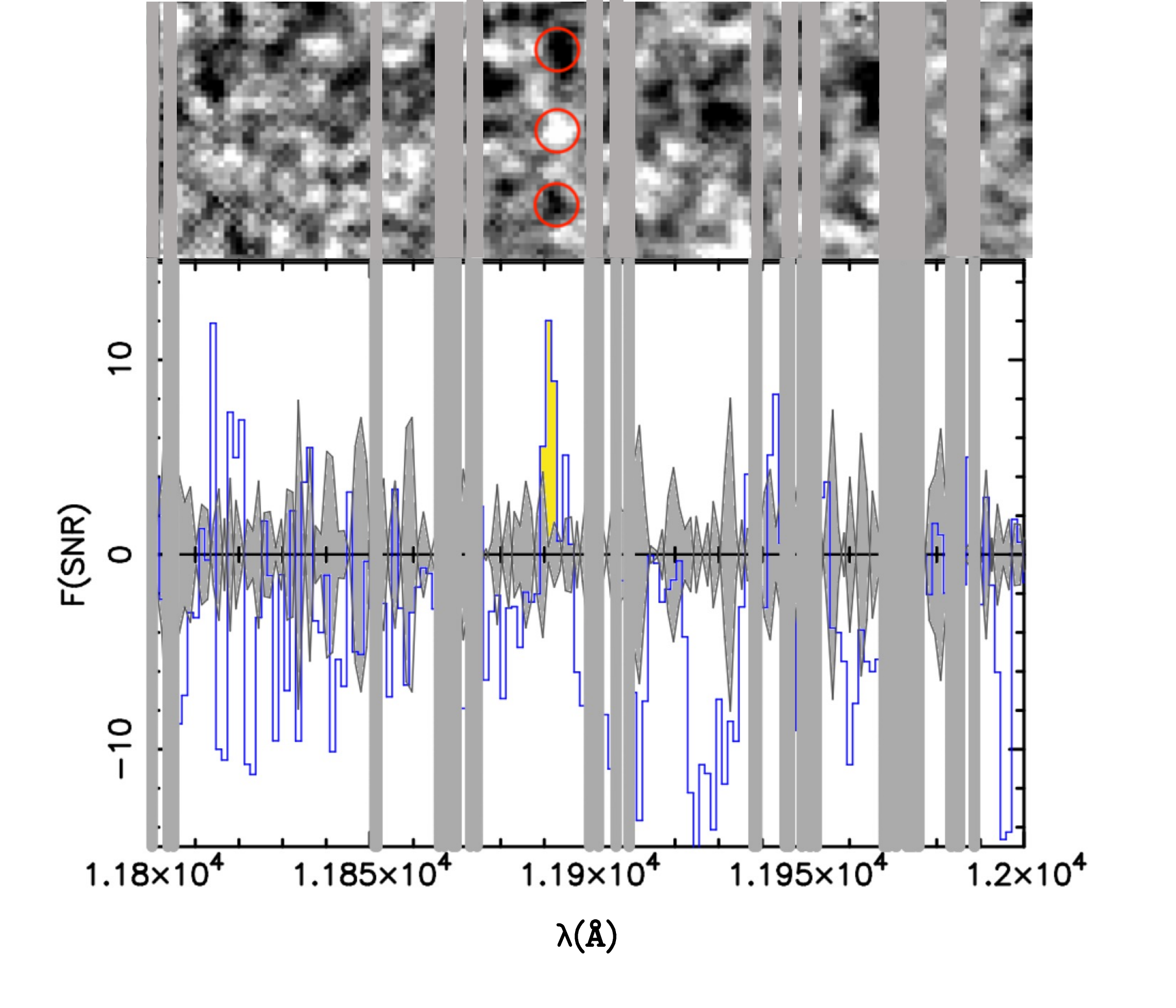}
    \caption{\label{GN-z10-3} (\textit{Top}) 2D MOSFIRE spectrum of GN-z10-3. Red circles show the location of a 5$\sigma$ emission line detected at $\lambda$=11891.4 corresponding to Lyman-$\alpha$ at $z$=8.78 in a pattern consistent with the telescope nodding. (\textit{Bottom}) 1D extracted spectrum (blue) with the emission line highlighted in yellow and the 1$\sigma$ noise level in grey. Grey rectangles indicate the position of sky lines.   }
\end{figure}

We observed GN-z10-3 with MOSFIRE/Keck undertaking a 5hrs 20min on source exposure in April 2019 (Keck 2019A\_U128, PI: B.E. Robertson). In order to monitor the seeing and transparency, two stars were included in the MOSFIRE mask. For the final analysis, only frames with seeing less than 0.8 arcsec FWHM were selected representing a total exposure of 4.9hrs. Data reduction was undertaken using the 2018 version of the MOSFIRE DRP. As the target was centred in the MOSFIRE slit, emission lines will feature a positive signal accompanied by two negative counterparts separated by the nodding scale of 1.25 arcsec. We identified a convincing ($>5\sigma$) emission line at $\lambda$=11891.4$\pm$1.3\AA\ with a FWHM=133$\pm$36km/s and an integrated flux of (1.37$\pm$0.24)$\times$10$^{-18}$cgs (Figure \ref{GN-z10-3}). No other line is present in the entire J-band spectra. Assuming this line is Lyman-$\alpha$, it corresponds to a redshift of $z$=8.78. In order to verify its reliability, we examined data comprising independent halves of the total exposure time. The line was retrieved with suitably reduced signal/noise on both spectra. The line width is comparable with that measured for other $z>7$ Lyman-$\alpha$ detections with MOSFIRE (e.g., \citealt{Zitrin2015}, \citealt{Finkelstein2013}, \citealt{Hoag2017}).

 The spectroscopic data refines the redshift value within the photometric redshift likelihood distribution (Figure \ref{fig.sed}). Although the IRAC 4.5 $\mu$m excess at this improved redshift below $z=9$ implies a likely contribution from [\OIII] 5007 \AA\ emission, we will return to the analysis of its SED in Section 5. Further sub-mm observations are currently on-going with NOEMA to detect [\OI] 145 $\mu$m, one of the brightest FIR emission lines \citep{Katz2019}. 

\begin{figure}
    \centering
        \includegraphics[width=8cm]{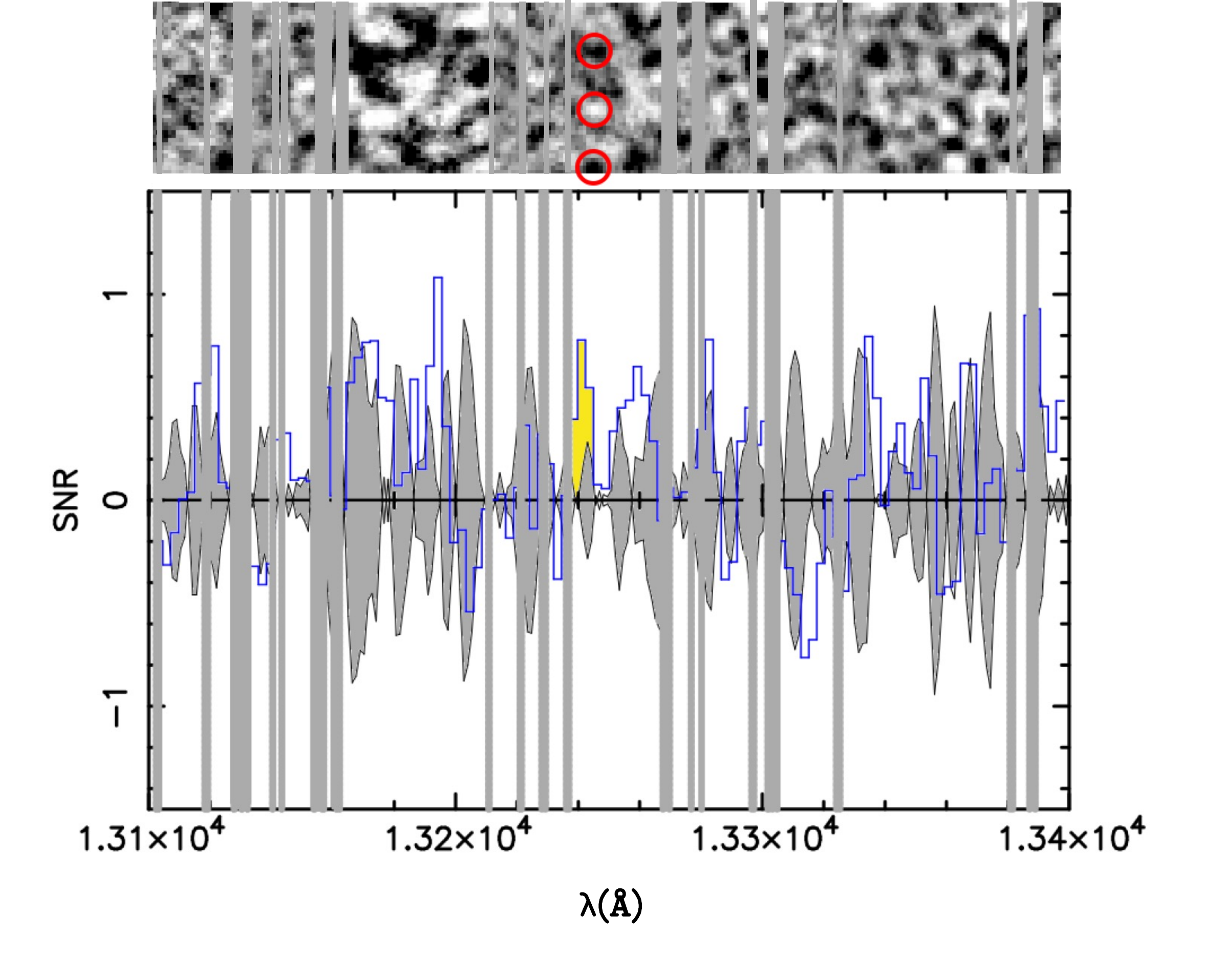}
    \caption{\label{GN-z9-1} (\textit{Top}) 2D MOSFIRE spectrum of GN-z9-1. Illustrated features follow those in Figure \ref{GN-z10-3}). (\textit{Bottom}) The 1D extracted spectrum in a 1.0 arcsec aperture is shown in blue. The grey line displays the 1$\sigma$ error. A tentative (3$\sigma$) emission line is highlighted in yellow which may be Lyman-$\alpha$ at $z$=9.89.   }
\end{figure}

During the same observing run, a separate mask was designed for GN-z9-1. A total on source exposure time of 5.6hrs was secured in good seeing conditions. The data was reduced and analysed in the manner described above. However, in this case, no clear ($>$5$\sigma$) emission line over the entire J-band. Reduced the significance threshold to 3$\sigma$, a faint feature is present at $\lambda$=13241.2\AA\ with a positive signal at the expected position of the target in the slit and two negative counterparts on each side. We estimated a flux of (2.2$\pm$0.7)$\times$10$^{-18}$cgs. Assuming this line is Lyman-$\alpha$, the redshift of GN-z9-1 would be $z=$9.89 (Figure~\ref{GN-z9-1}). However, although consistent with the photometric redshift likelihood distribution (Figure \ref{fig.sed}), we consider this a marginal claim. Deeper spectroscopic observation, such as those that will become feasible with JWST, are needed to confirm this redshift.

\subsection{UVISTA-1212}

UVISTA-1212, one of the brightest $z\geq$8 candidates known to date, was observed with FLAMINGOS-2 on the Gemini-South telescope via the Fast Turnaround programme (ID: GS-2020A-FT-102, PI: Roberts-Borsani). Observations were accomplished in service mode in February 2020 in J-band using the R3K grism and a 4 pixel wide slit. From an allocation of 9.8 hours, a 8.8 hour on source exposure was taken in good seeing conditions.

\begin{figure}
    \centering
        \includegraphics[width=8cm]{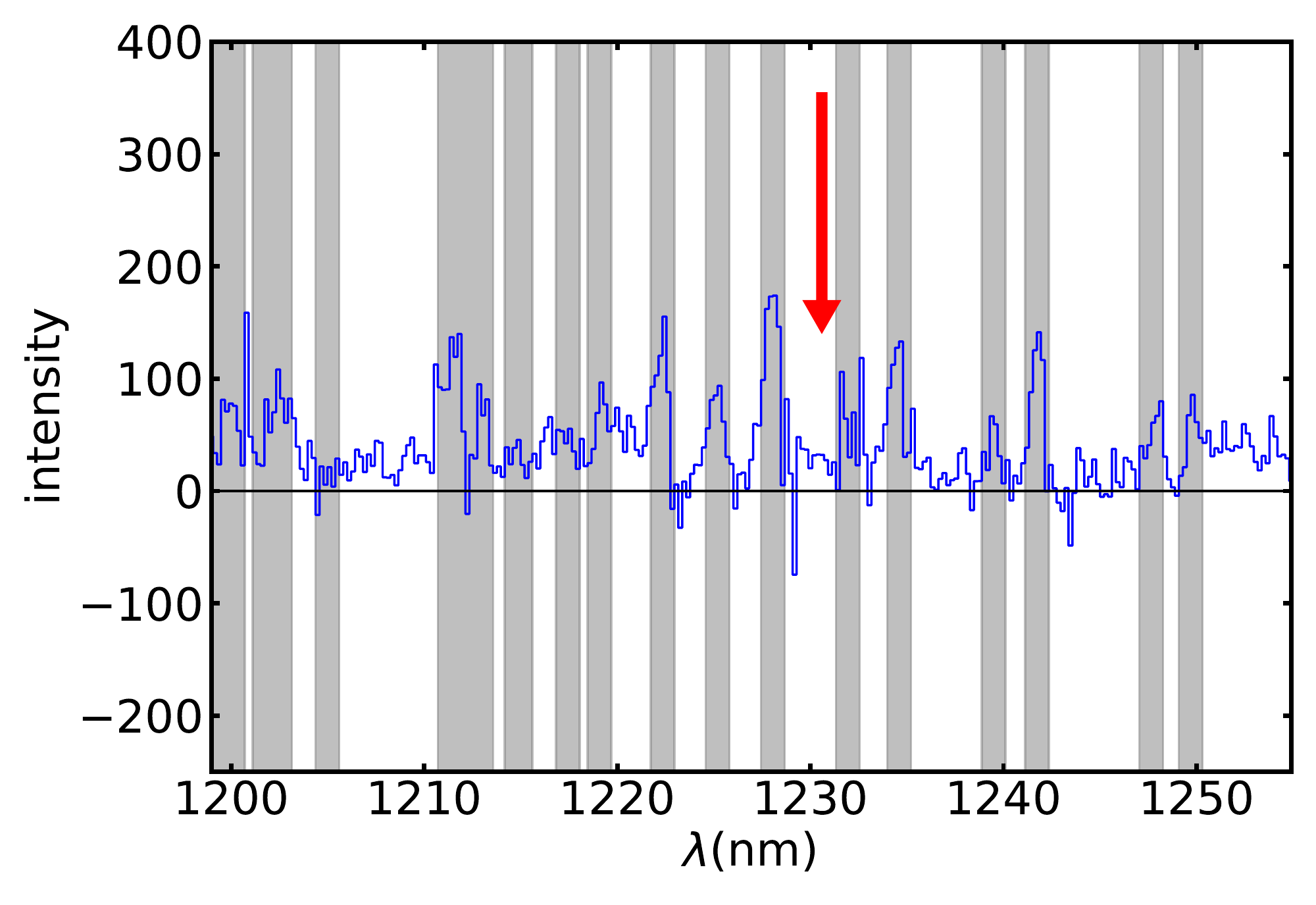}
    \caption{\label{f2_1dspec} The portion of the FLAMINGOS2 spectrum where Lyman-$\alpha$ may lie according to the photometric redshift uncertainties for UVISTA-1212. No convincing emission line is found. Grey rectangles show the position of sky lines.}
\end{figure}

The data were reduced using the FLAMINGOS-2 Python cookbook for longslit observations (specifically the \textit{reduce\_ls} script, which primarily uses the PyRaf module as a wrapper around IRAF functions to reduce the data). More specifically, the \textit{nsflat}, \textit{nsreduce} and \textit{nscombine} were used to reduce the dark, flat and arc frames, as well as the standard star and science observations. The final 1D spectrum is shown in Figure \ref{f2_1dspec}. After a careful inspection of the reduced spectra, no clear emission line was found over the entire J-band spectrum.

\subsection{GS-z9-1}

\begin{figure}
    \centering
        \includegraphics[width=9cm]{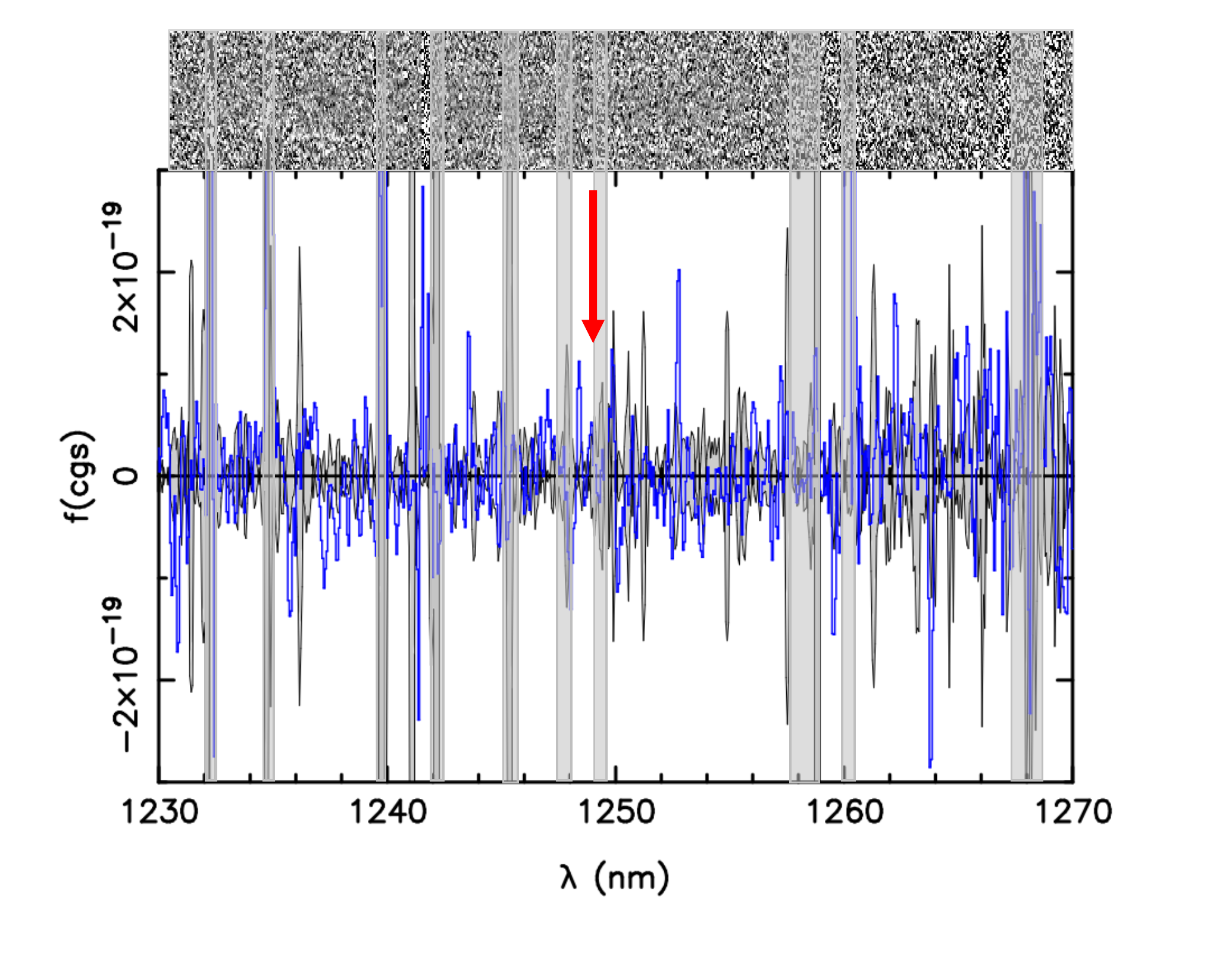}
    \caption{\label{fig:gs-z9-2_spec} The portion of the X-Shooter spectrum where Lyman-$\alpha$ may lie according to the photometric redshift uncertainties for GS-z9-1. No convincing emission line is found. The blue line shows the 1D extracted spectrum at the position of the candidate, grey regions show the 1$\sigma$ noise extracted in the same sized aperture, and light blue rectangles show the position of sky lines. The red arrow displays the position of Ly-$\alpha$ at $z_{phot}$=9.26}.
\end{figure}

GS-z9-1 has a photometric redshift of 9.26$^{+0.41}_{-0.42}$ and we have obtained 9.8 hours of observing time (8.0 hours on source) from an allocation of 13.5 hours with X-Shooter/VLT (ID : 106.20ZH.001, PI: R. Ellis). Observations were undertaken in service mode in good seeing conditions ($<$1.1 arcsec) between October 2020 and January 2021 \footnote{We thank ESO staff for observing during both Christmas and New Year.} We carefully checked the centring of the source in the X-Shooter slit and applied the same data reduction procedure as described in section \ref{sec.macs0416}. Unfortunately no clear emission line is seen after a first inspection of the data (Figure \ref{fig:gs-z9-2_spec}). At the expected position of Ly-$\alpha$ at $z_{phot}$, we measure a 1$\sigma$ upper limit of 7$\times$10$^{-19}$cgs.

%{\it RSE - consistent with Figures 3, 6, I think we should show the spectrum and absence of any feature at the likely photo-z.}

\section{Physical Properties}
\label{sec.prop}

\subsection{Updating the Properties}
\label{sec.ages}

We now reconsider the physical properties of our sample, taking into account, where available, the newly-acquired spectroscopic redshifts (Table \ref{tab.spectro}). In addition to MACS1149-JD1, we now have additional redshifts for MAC0416-JD and GN-z10-3, representing half of our original sample. Although we have marginal evidence that GN-z9-1 lies beyond $z\simeq$9, for it, GS-z9-1 and UVISTA-1212, we will continue to assume the photometric redshifts given in Table \ref{list}. A re-analysis of the SED is particularly relevant for GN-z10-3 whose spectroscopic redshift is now confirmed to lie below $z=9$, implying a possible significant contribution to the IRAC excess from rest-frame optical emission lines. The updated results based on BAGPIPES fits to the SEDs, now constrained with a fixed spectroscopic redshift, are given in Table \ref{fig:prop.spec}. In the three cases where the spectroscopic redshift is confirmed to be above $z\simeq$9, the derived properties are naturally similar to those given in Table \ref{list}. With the exception of UVISTA-1212, which appears to be an outlier in our sample, all galaxies have a stellar mass ranging from $\sim$ 1 to 4$\times$10$^9$ M$_{\odot}$, a star formation rate ranging from $\sim$10 to 30 M$_{\odot}$/yr and a low dust attenuation ($<$0.5 mag). However, for GN-z10-3, despite a spectroscopic constraint of $z$=8.78, a re-analysis of the SED incorporating this value may still be consistent with a mature stellar population. This may appear surprising given the IRAC excess could be readily explained with strong [\OIII] 5007 \AA\ emission. 

\begin{figure}
    \centering
        \includegraphics[width=9cm]{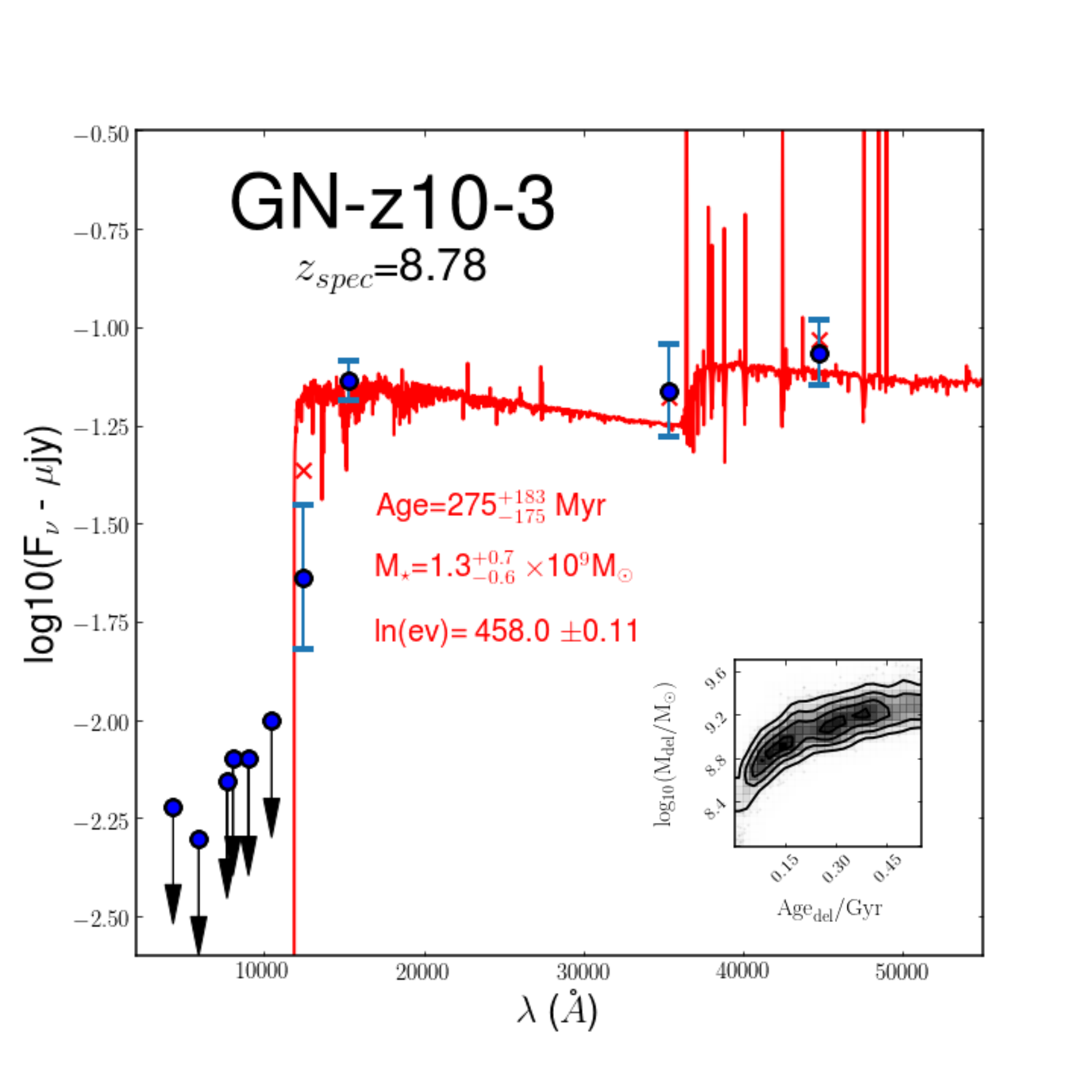}
    \caption{\label{GN-z10-3REV} Best fit to the spectral energy distribution of GN-z10-3 now constrained at the spectroscopic redshift of $z=$8.78. Blue points and limits represent the observed photometry and red crosses show the expected measurements. The posterior distribution of the age and stellar mass is now bimodal, indicating it is not possible to distinguish the relative contributions of [O III] emission and a Balmer break to the IRAC excess (see text).} 
\end{figure}

 Figure \ref{GN-z10-3REV} shows a re-analysis of the SED shown in Figure \ref{fig.sed} with the redshift appropriately constrained at $z=8.78$. The posterior distribution in age and stellar masses is now bimodal and thus it is not possible to  distinguish between a young population whose [O III] 5007 \AA\ emission contributes to the IRAC excess, and an older solution with a contribution from a Balmer break. Similar interpretational ambiguities were seen for $z<9$ sources with IRAC excesses by \citet{Roberts-Borsani2020} who argued this degeneracy may be broken with independent measures of [O III] 88$\mu$m emission. In the case of the young solution for GN-z10-3, the required equivalent width (EW) of [\OIII]+H$\beta$ is $\sim$1000 \AA\ which is consistent with the range observed in some high redshift galaxies (e.g., \citealt{Strait2020}, \citealt{Bowler2020}). The flat continuum between F160W and 360$\mu$m would then imply that [O II] has an EW $> 50 $\AA. Recent studies suggest this is close to the maximum possible for a young starburst (e.g. \citealt{Yang2017}). Although we are unable to distinguish between these two fits with the current data, we adopt the properties of an older stellar population in Table \ref{fig:prop.spec}.

\subsection{The $z\sim$9 Lyman-$\alpha$ fraction}

We now consider the question of the visibility of Lyman-$\alpha$ deep in the reionisation era. This can be done in terms of a Lyman-$\alpha$ emission fraction for our sample. Thus we have two targets with a secure Lyman-$\alpha$ detection and one (GN-z9-1) which is tentative (see Section \ref{sec.spectro}). All candidates have reliable photometric redshifts at $z\sim9$ or a far infrared emission line, suggesting the likelihood of contamination of the sample by low redshift interlopers is small. Despite obvious uncertainties for such a small sample, the Lyman-$\alpha$ emission fraction calculated for the sample of $6$ $z\sim 9$ galaxies presented in this paper is very high: $\chi_{\rm{LAE}} = 0.33_{-0.21}^{+0.28}$ if we include GN-z9-1 or $\chi_{\rm{LAE}} = 0.50^{+0.26}_{-0.26}$ otherwise.

\begin{figure}
    \centering
    \includegraphics[width=0.48\textwidth]{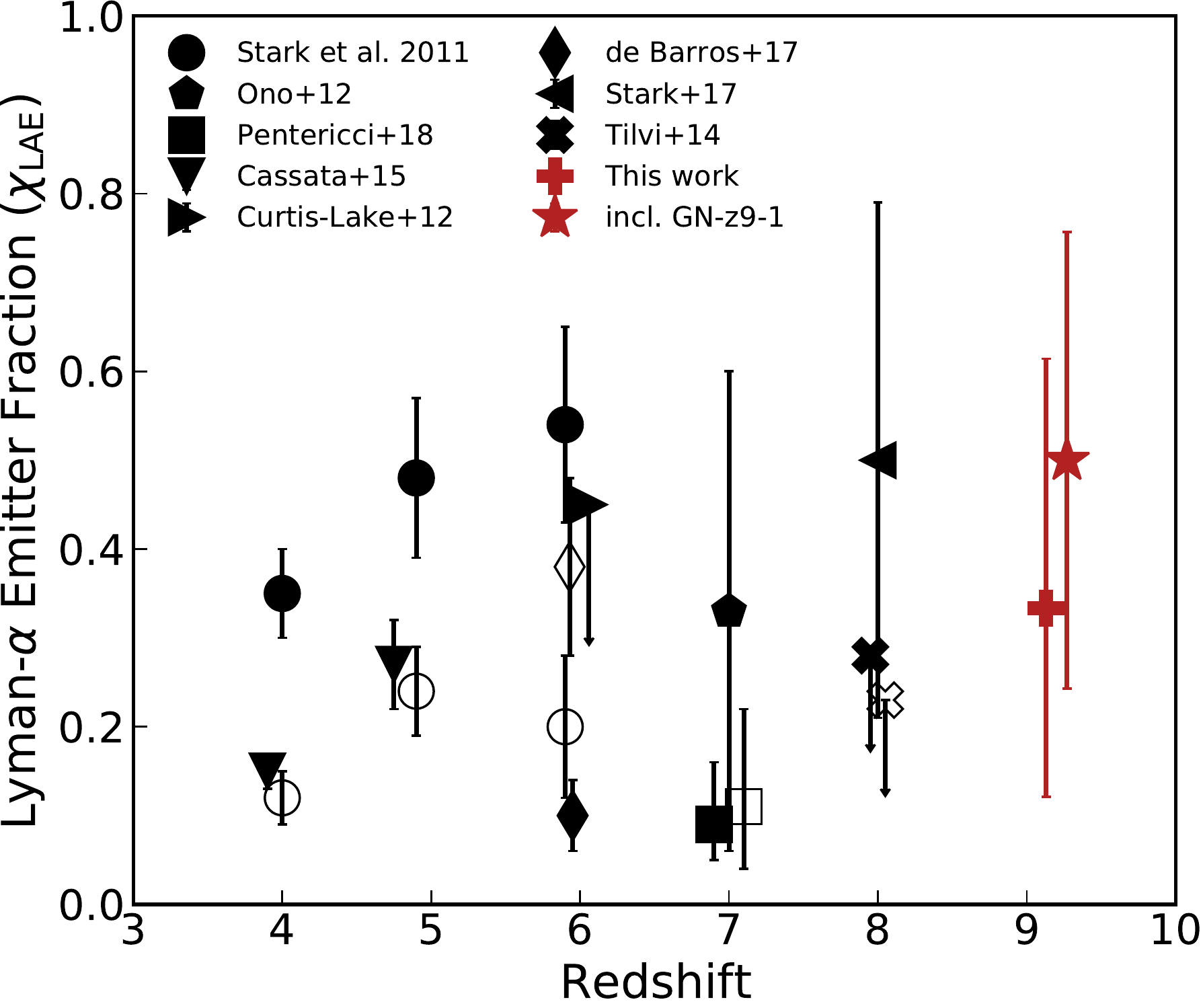}
    \caption{Fraction of Lyman-break galaxies revealing Lyman-$\alpha$ emission with an equivalent width $\rm{EW}>25$ \AA. Full symbols represent the more luminous population with $-21.75<M_{\rm{UV}} < -20.25$ and the empty symbols indicate galaxies with $ -20.25<M_{\rm{UV}} < -18.75$. The fraction for galaxies in the present sample is indicated by the red symbols.}
    \label{fig.Lya_fraction}
\end{figure}

This high fraction of detections continues a trend of recent results (\citealt{Roberts-Borsani2016,Stark2017,Mason2018}) that have weakened the originally-claimed rapid decline above $z\simeq$7 (\citealt{Pentericci2011,Schenker2012,Schenker2014}). Our results are compared with these earlier trends in Figure \ref{fig.Lya_fraction}). As with \citet{Roberts-Borsani2016} and \citet{Stark2017}, we explicitly selected bright galaxies amenable to ground-based spectroscopy with red \textit{Spitzer}/IRAC colours (3.6$\mu$m - 4.5$\mu$m > 0.5, Section \ref{sec.selection}). However, while the 100\% visibility of Lyman-$\alpha$ in \citet{Roberts-Borsani2016}'s sample might be attributed to a correspondingly strong rest-frame optical [\OIII] emission responsible for the IRAC excess, for the three $z>9$ targets in the present sample, only a Balmer break synonymous with a mature stellar population (e.g., \citealt{Hashimoto2018}) could be responsible. This suggests that the enhanced Lyman-$\alpha$ fraction arises from the fact that some luminous galaxies, regardless of their [\OIII] emission, are capable of producing early ionised bubbles. As discussed by \citet{Mason2018}, much of the early work probing the redshift-dependent fraction necessarily targeted sub-luminous galaxies in order to efficiently exploit multi-slit spectrographs with limited fields of view.

There is good evidence that intense [\OIII] emitters at low and intermediate redshifts have high LyC escape fractions \citep{Izotov2018,Fletcher2019,Nakajima2020} and/or ionising efficiencies \citep[e.g.,][]{Nakajima2016}, consistent with the creation of ionised bubbles at high-redshift. What remains unclear is how $z>9$ galaxies with IRAC excesses indicating a mature stellar population can emit enough Lyman continuum photons to ionise large bubbles. The Str\"omgren radius of the ionised bubble created by a galaxy scales with $\propto (t_{\rm{em}}f_{\rm{esc,LyC}})^{1/3}$ \citep[e.g.,][]{Cen2000}, where $f_{\rm{esc,LyC}}$ is the escape fraction of Lyman continuum photons and $t_{\rm{em}}$ the time the galaxies has been emitting such photons. 

Assuming an average ionising efficiency and $f_{\rm{esc,LyC}} < 3\%$, the galaxies in our $z\sim 9$ sample, which all sustained star formation over $\gtrsim 100-200$ Myr (see Table \ref{list}), can create an ionised bubble as large as those of COLA1 \citep{Hu2016,Matthee2018} or A370p\_z1 \citep{Meyer2021}, two actively star forming $z>6$ galaxies with very high escape fractions. Such a large ionised bubble may explain why Lyman-$\alpha$ is observed in MACS1149-JD1 blueshifted by $\sim -450 \text{km s}^{-1}$ with respect to [\OIII] 88 $\mu$m. In summary therefore, the high detected fraction of Lyman-$\alpha$ emission at $z\simeq$9 in luminous galaxies may offer further support of their mature stellar populations.

\subsection{How Representative is our Sample?}
Early evidence for mature stellar populations at $z>7.5$ \citep{Hashimoto2018,Roberts-Borsani2020} has been hard to reconcile with numerical simulations of the first galaxies. While \citet{Binggeli2019} conclude that $z\sim9$ Balmer breaks must be very rare, \citealt{Katz2019} find in the contrary that they can be formed easily in their simulations, although dust might complicate the interpretation. In light of our results, it is thus important to assess observationally the prevalence of $z\sim 9$ galaxies with older stellar populations. 

While we cannot compare Balmer-break objects with a larger population of spectroscopically confirmed $z\sim9$ galaxies, we can assess the prevalence of Balmer-break objects in a parent sample of $z>9$ photometric candidates. In order to investigate the relative abundances of the two populations, we have carefully selected $z>9$ galaxies candidates using the same procedure and magnitude cuts as in Section \ref{sec.selection}, but removing the IRAC excess criterion. The candidates were visually inspected by two authors (NL, RAM). Due to the absence of filters bluewards of $<8500$\, \AA\, in ULTRAVISTA, we restricted our search to the Frontier Fields and CANDELS. This additional search results in $14$ photometric candidates (including the $5$ Balmer-break objects presented above). Balmer-break objects thus appear to represent a  significant fraction($\simeq 0.36^{+0.17}_{-0.14}$)  of the $z>9$ bright galaxy population, in contrast to the results of numerical simulations. Indeed this Balmer-break fraction could represent a lower limit since the denominator comprises galaxies without spectroscopic redshifts: several could be low redshift interlopers or $8<z<9$ galaxies (given the large photometric redshift errors) whose IRAC excess may not arise from a Balmer Break. This is also the case for the nominator (e.g. GS-z9-1), although to a lower extent.

\begin{figure}
    \centering
    \includegraphics[width=0.45\textwidth]{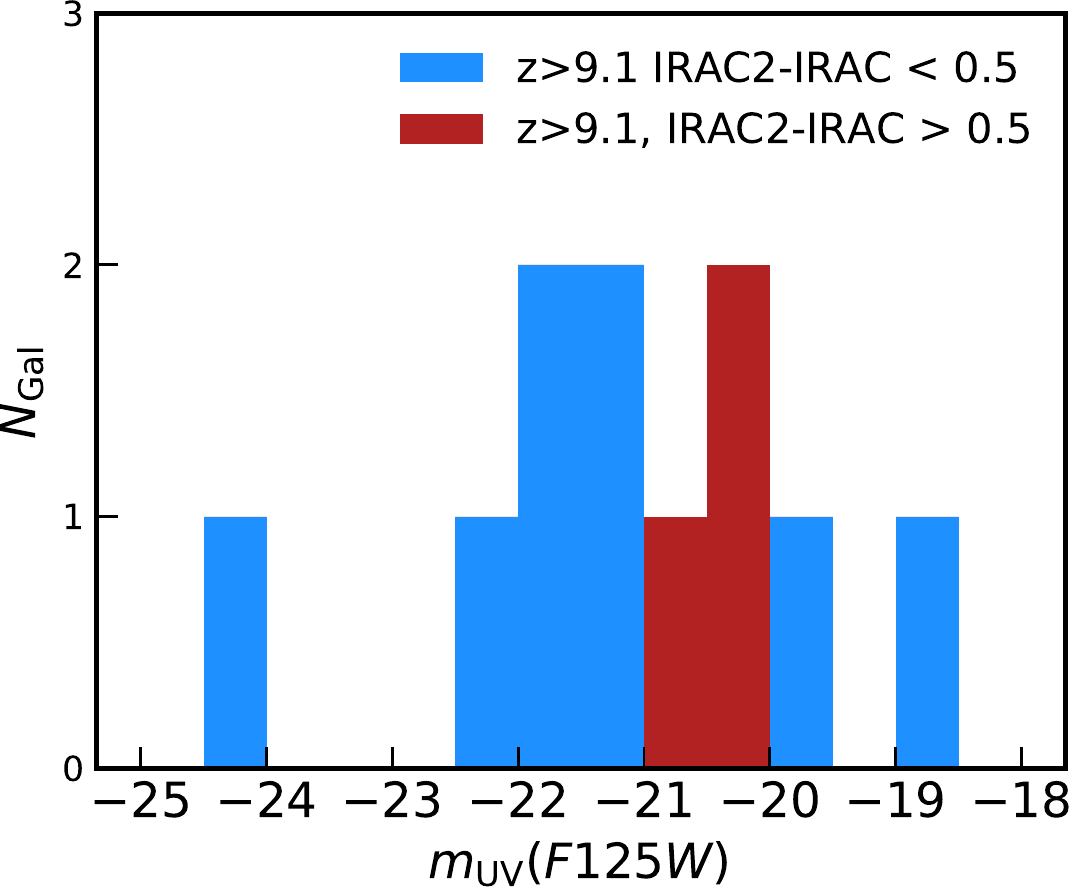}
    \caption{A comparison of the rest-frame UV  magnitudes for $z\sim9$ candidates in CANDELS selected without regard to their IRAC excesses with those for the three GOODS-S/N sources discussed in this article.}
    \label{fig:comparison}
\end{figure}

We can compare the rest-frame UV luminosities of this parent sample with those of our Balmer break galaxies. As most of the candidates are drawn from the GOODS-North/South  fields($11/14$), we restrict the comparison to the CANDELS fields, ensuring we can avoid the need to make corrections for lensing magnification in the cluster fields. We derive the rest-frame UV magnitude of our candidates using the F125W magnitude, the best-fit photometric redshift and a UV slope $\alpha=2, (f_\nu \propto \nu^\alpha)$ for the K-correction. We compare the rest-frame UV magnitudes for the two samples in Figure \ref{fig:comparison}. Although the statistics are poor, our objects with red IRAC colors are not notably brighter or fainter than blue IRAC color objects. A two-sided KS test is inconclusive and the null hypothesis that the two distributions are different cannot be rejected ($p=0.12$).  Thus, given the modest sample size to date, our $z>9$ Balmer break sources remain broadly representative of the larger $z>9$ galaxy population. We note that all $z>9$ candidates represent the brightest objects existing at that early time due to obvious limited depth of the imaging data. The hypothesis that Balmer break objects are less frequent in the fainter population is plausible, but can only be tested with \textit{JWST}.

\section{Discussion}
\label{sec.discussion}

In this article, we have argued that from a sample of six $z\simeq$9 galaxies with \textit{Spitzer}/IRAC excess for which spectroscopic redshifts have now been secured for three, possibly four, we infer past star formation histories consistent with significant contributions beyond a redshift $z=10$.

\begin{figure}
%      \hspace{-1.75cm}
      \centering
  \includegraphics[width=1.0\columnwidth]{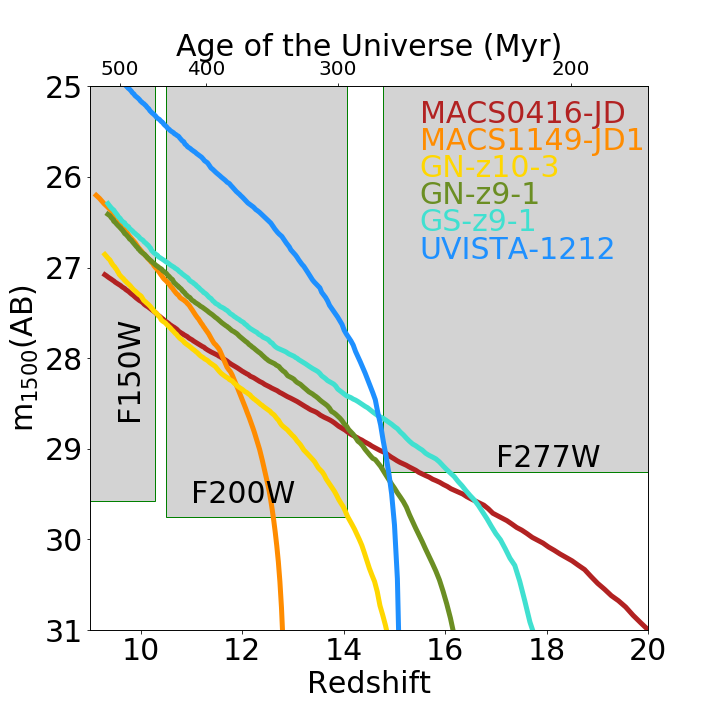}
    \caption{Earlier redshift evolution of the rest-frame 1500\AA\ luminosity computed from our SED fits for the six galaxies in our sample. %With the exception of GN-z10-3, the precursors are expected to be comparably luminous or more luminous at earlier times. 
    Filled regions show the 5$\sigma$ sensitivity of NIRCam/JWST filters (F150W, F200W and F277W) in 3hrs. }
    \label{fig:past_luminosity}
\end{figure}

%Following the approach adopted by \citet{Hashimoto2018} and \citet{Roberts-Borsani2020}, 
We can estimate the earlier rest-frame UV luminosity for each of the six galaxies in our sample assuming the best-fit star formation history (SFH). As described in Section \ref{sec.photometry}, the best SED fit is obtained for most of the objects with an evolved stellar population. Table \ref{list} shows the best-fit parameters for all galaxies, updated in Table \ref{fig:prop.spec} with the spectroscopic redshift for 3 of them. Based on these SFHs  we compute the rest-frame $\rm{L_{1500}}$ luminosity as a function of redshift from $z\sim$15 to the observed redshift taking into account the dust reddening. For those galaxies without spectroscopic redshifts, the photometric value is assumed. Figure \ref{fig:past_luminosity} shows the evolution of $\rm{m_{1500}}$ as a function of redshift and demonstrates that each galaxy is as luminous in its earlier life. While these past star formation histories are clearly illustrative, they serve to demonstrate the feasibility of searches for earlier progenitors with JWST (e.g., \citealt{Roberts-Borsani2021}).

\begin{figure}
%      \hspace{-1.75cm}
      \centering
 \includegraphics[width=1.0\columnwidth]{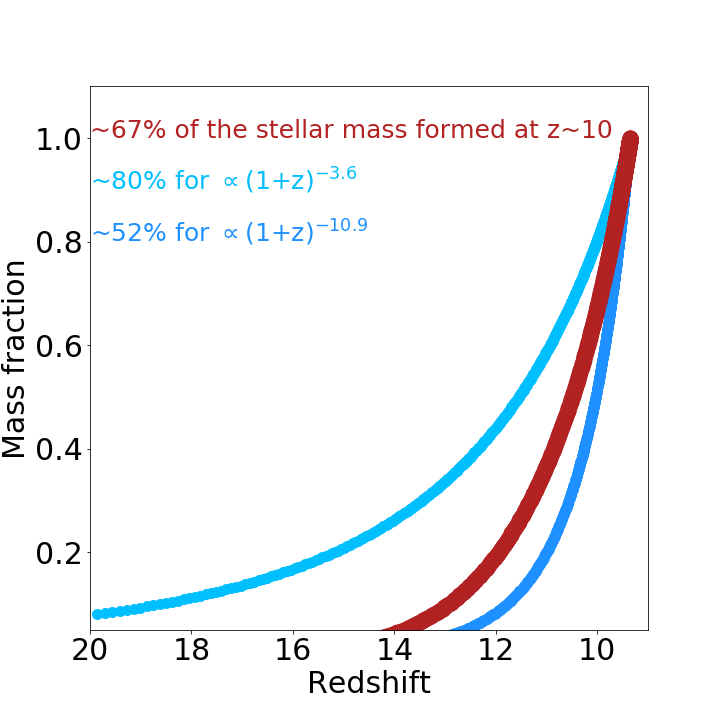}
    \caption{Redshift evolution of the assembled stellar mass computed from the SFH deduced from our SED fits and averaged over the 6 galaxies discussed in this paper (red). This is compared with that predicted for two contrasting measures of the rate of decline in the UV luminosity density deduced from large photometric surveys (\citealt{Oesch2014}, \citealt{McLeod2016}). We estimate that $\simeq$ 70\% of the stellar mass was already in place by $z=10$ for the 6 galaxies in our $z\geq$9 sample.} 
    \label{fig:building_mass}
\end{figure}

From our SFHs, we can also determine the redshift-dependent history of the assembled stellar mass providing independent insight into the evolution of the UV luminosity density within the first 500 Myr (Figure \ref{fig:building_mass}). Two contrasting hypotheses are still debated in the literature: a rapid \citep{Oesch2018} or a smooth \citep{McLeod2016} decline of the UV luminosity density over $8 < z < 11$. For the case of a fast decline ($\propto$(1+$z$)$^{-11}$), the luminous galaxies probed would have formed $\sim$50\% of their stellar mass by redshift $z=$10, whereas for the smooth decline case ($\propto$(1+$z)^{-4}$) $\sim$80\% of the stellar mass is formed at $z=$10. Although our sample is admittedly small, it represents the only relevant data currently available. Averaging the SFHs of our 6 galaxies we determine that they formed $\sim$70\% of their mass by $z=$10, which seems to favour a smooth decline.

\begin{table*}
\hspace{-1.4cm} 
    \centering    
    \hspace{-1.4cm}  \begin{tabular}{l|cccc} \hline
Target & $z_{\text{spec}}$ &  $\rm{M_{\star}}$[$\times$10$^9$M$_{\odot}$] &  Age (Myr) & $f_M$(z$>$10) [\%]\\ \hline
MACS0416-JD & 9.28 & 1.37$^{+0.75}_{-0.55}$ &  351$^{+115}_{-153}$  & 71.8$^{+10.4}_{-20.3}$\\
MACS1149-JD1 & 9.11 & 0.66$^{+0.09}_{-0.04}$ &  192$^{+159}_{-87}$   & 47.5$^{+20.0}_{-14.2}$  \\
GN-z10-3 &  8.78 &  1.27$^{+0.70}_{-0.61}$ &  275$^{+184}_{-175}$  & 39.1$^{+24.6}_{-38.9}$\\
 \hline
\end{tabular}
\caption{\label{fig:prop.spec}  Updated properties for those sources in Table \ref{list} with a spectroscopic redshift. The final column displays the percentage of the presently-observed stellar mass already formed before $z=$10.}

\end{table*}

%\section{Conclusions}
\section*{Data Availability}
The data underlying this article will be shared on reasonable request to the corresponding author.

\section*{Acknowledgements}
NL acknowledges support from the Kavli Foundation.
RAM and RSE acknowledge funding from the European Research Council (ERC) under the European Union's Horizon 2020 research and innovation programme (grant agreement No 669253). BER was supported in part by NASA program HST-GO-14747, contract NNG16PJ25C, and grant 80NSSC18K0563, and NSF award 1828315. We thank Adam Carnall for providing an updated version of BAGPIPES allowing high ionisation parameter.

Based on observations made with ESO Telescopes at the Paranal Observatory under programme ID 0104.A-0028. This paper makes use of the following ALMA data: ADS/JAO.ALMA\#2019.1.00061.S. ALMA is a partnership of ESO (representing its member states), NSF (USA) and NINS (Japan), together with NRC (Canada), MOST and ASIAA (Taiwan), and KASI (Republic of Korea), in cooperation with the Republic of Chile. The Joint ALMA Observatory is operated by ESO, AUI/NRAO and NAOJ. Based on observations obtained at the international Gemini Observatory, a program of NSF’s NOIRLab, which is managed by the Association of Universities for Research in Astronomy (AURA) under a cooperative agreement with the National Science Foundation on behalf of the Gemini Observatory partnership: the National Science Foundation (United States), National Research Council (Canada), Agencia Nacional de Investigación y Desarrollo (Chile), Ministerio de Ciencia, Tecnología e Innovación (Argentina), Ministério da Ciência, Tecnologia, Inovações e Comunicações (Brazil), and Korea Astronomy and Space Science Institute (Republic of Korea). The data presented herein were obtained at the W. M. Keck Observatory, which is operated as a scientific partnership among the California Institute of Technology, the University of California and the National Aeronautics and Space Administration. The Observatory was made possible by the generous financial support of the W. M. Keck Foundation. The authors wish to recognize and acknowledge the very significant cultural role and reverence that the summit of Maunakea has always had within the indigenous Hawaiian community.  We are most fortunate to have the opportunity to conduct observations from this mountain. ID: 2019

%%%%%%%%%%%%%%%%%%%%%%%%%%%%%%%%%%%%%%%%%%%%%%%%%%



\bibliographystyle{mnras}
\bibliography{example} % if your bibtex file is called example.bib

% Don't change these lines
\bsp	% typesetting comment
\label{lastpage}
\end{document}